\documentclass[
 aps,pra,
 superscriptaddress,
 amsmath,
 amssymb,
 reprint,
 showkeys
]{revtex4-1}

   \usepackage{graphicx}
   \usepackage{dcolumn}
   \usepackage{bm}
   \usepackage{color}
   \usepackage{titlesec}
   \usepackage{tabularx}
   \usepackage{array}

   \usepackage{comment}

   \usepackage[normalem]{ulem}
   \usepackage{longtable}
   \usepackage{multirow}
   
   \usepackage[T1]{fontenc}
   \usepackage[utf8]{inputenc}
   \usepackage[english]{babel}
   
   \usepackage[cmyk,dvipsnames]{xcolor}

   \usepackage{braket}

   \makeatletter
\renewcommand*\env@matrix[1][\arraystretch]{%
  \edef\arraystretch{#1}%
  \hskip -\arraycolsep
  \let\@ifnextchar\new@ifnextchar
  \array{*\c@MaxMatrixCols c}}
\makeatother

\begin{document}

\title{Toward room-temperature nanoscale skyrmions in ultrathin films}

\author{Anastasiia S. Varentcova}
\affiliation{ITMO University, 197101 St. Petersburg, Russia}
\affiliation{Science Institute of the University of Iceland, 107 Reykjav\'ik, Iceland}
\affiliation{Kavli Institute of Nanoscience, Delft University of Technology, P.O. Box 4056, 2600 GA Delft, The Netherlands}

\author{Stephan von Malottki}
\affiliation{Institute of Theoretical Physics and Astrophysics, University of Kiel, Leibnizstrasse 15, 24098 Kiel, Germany}

\author{Maria N. Potkina}
\affiliation{ITMO University, 197101 St. Petersburg, Russia}
\affiliation{Science Institute of the University of Iceland, 107 Reykjav\'ik, Iceland}
\affiliation{St. Petersburg State University, 198504 St. Petersburg, Russia}

\author{Grzegorz Kwiatkowski}
\affiliation{Science Institute of the University of Iceland, 107 Reykjav\'ik, Iceland}

\author{Stefan Heinze}
\affiliation{Institute of Theoretical Physics and Astrophysics, University of Kiel, Leibnizstrasse 15, 24098 Kiel, Germany}

\author{Pavel F. Bessarab}
\email[Corresponding author. E-mail: ]{bessarab@hi.is}
\affiliation{ITMO University, 197101 St. Petersburg, Russia}
\affiliation{Science Institute of the University of Iceland, 107 Reykjav\'ik, Iceland}
\affiliation{Institute of Theoretical Physics and Astrophysics, University of Kiel, Leibnizstrasse 15, 24098 Kiel, Germany}
\affiliation{Peter Gr\"unberg Institute and Institute for Advanced Simulation, Forschungszentrum J\"ulich, 52425 J\"ulich, Germany}

\begin{abstract}
Breaking the dilemma between small size and room temperature stability is a necessary prerequisite for skyrmion-based information technology. Here we demonstrate by means of rate theory and an atomistic spin Hamiltonian that the stability of isolated skyrmions in ultrathin ferromagnetic films can be enhanced by the concerted variation of magnetic interactions while keeping the skyrmion size unchanged. We predict film systems where the lifetime of sub-10 nm skyrmions can reach years at ambient conditions. The long lifetime of such small skyrmions is due to exceptionally large Arrhenius pre-exponential and the stabilizing effect of the energy barrier is insignificant at room temperature. A dramatic increase in the pre-exponential is achieved thanks to softening of magnon modes of the skyrmion, thereby 
increasing the entropy of the skyrmion with respect to the transition state for collapse. Increasing the number of skyrmion deformation modes should be a guiding principle for the realization of nanoscale, room-temperature stable skyrmions.
\end{abstract}

\keywords{magnetic skyrmion; thermal stability; lifetime; energy barrier; entropy barrier; Arrhenius pre-exponential; skyrmion shape; deformation mode}

\maketitle

\section*{Introduction}
\noindent Encoding data with metastable magnetic skyrmions~\cite{Bogdanov1989,Bogdanov1994,bogdanov_1994} is an appealing solution for future information technology~\cite{kiselev_2011,fert_2013,nagaosa_2013}. 
Although potential feasibility of this concept has been demonstrated by experimental detection of skyrmions in various magnetic systems~\cite{woo_2016,moreau_2016,boulle_2016,yu_2016,jiang_2015,legrand_2017,chen_2015,hsu_2017,soumyanarayanan_2017,jiang_2017,hrabec_2017,litzius_2017,heinze_2011,romming_2013,romming_2015, hanneken_2015, caretta_2018,hsu_2018,meyer_2019}, 
realization of viable skyrmion-based digital devices is an ongoing, challenging problem. 
Ultimately, the size of the skyrmionic bits should not exceed 10 nm so as to 
improve on the data density level already achieved in conventional technology. For such small skyrmions, thermal stability becomes an issue as thermal fluctuations can induce spontaneous collapse of the skyrmion state and, therefore, corrupt the stored data. Indeed, sub-10 nm skyrmions have so far been detected 
only at very low temperatures~\cite{heinze_2011,romming_2013,romming_2015,hanneken_2015,hsu_2017,hsu_2018,meyer_2019}.  
On the other hand, room temperature stability has been reported for larger skyrmions~\cite{chen_2015,woo_2016,moreau_2016,boulle_2016,yu_2016,jiang_2015,legrand_2017,soumyanarayanan_2017,jiang_2017,hrabec_2017,litzius_2017,caretta_2018}. 
The challenge is to design materials where skyrmions are both small and long-lived at ambient conditions.  

\indent 
Theoretical calculations can help engineer skyrmions towards applications by evaluating skyrmion lifetime as a function of material parameters, applied stimuli, and temperature. 
The lifetime $\tau$, i.e. the mean time it takes the magnetic system coupled to the heat reservoir of temperature $T$ to escape from the energy well corresponding to the metastable skyrmion state, can be 
described by the Arrhenius expression:
\begin{equation}
    \tau = \tau_0\exp\left(\frac{\Delta E}{k_\text{B} T}\right),
    \label{eq:arrhenius}
\end{equation}
and characterized by 
the annihilation energy barrier $\Delta E$ and the pre-exponential $\tau_0$.  
Previous theoretical studies have focused on calculations of the energy barrier with respect to the radially symmetric skyrmion collapse~\cite{bessarab_2015,lobanov_2016,malottki_2017} 
and skyrmion escape through the boundary of the system~\cite{stosic_2017,uzdin_2017,cortes_2017}. Asymmetric collapse of a skyrmion in systems with frustrated magnetic exchange interaction has also been reported in recent studies~\cite{meyer_2019,desplat_2019,heil_2019}. 
Dependencies of $\Delta E$ on various magnetic interaction parameters~\cite{stosic_2017,varentsova_2018}, 
external magnetic field~\cite{bessarab_2018,uzdin_2017b}, defects~\cite{uzdin_2017,stosic_2017b}, and frustration in the magnetic exchange~\cite{malottki_2017} have previously been identified by means of minimum energy path calculations within atomistic spin models. 
Recently, B\"uttner {\it et al.}~\cite{buttner_2018} have undertaken an extensive analysis of the phase diagram for isolated skyrmions within the continuous magnetization framework, where the annihilation energy barrier was approximated by the universal energy of the zero-diameter skyrmion relative to the skyrmion energy minimum and the pre-exponential factor was treated as a phenomenological constant. The study 
concluded that ultrasmall skyrmions in materials exhibiting local ferromagnetic order of spins could not be stable at ambient conditions~\cite{buttner_2018}.

However, knowledge about the energy barrier alone does not provide a complete picture of the thermal stability. Entropic and dynamical effects must also be evaluated for the prediction of the lifetime. 
These contributions are incorporated in the pre-exponential factor $\tau_0$, inverse of which is often referred to as the attempt frequency. 
Definite evaluation of the prefactor appears to be particularly important for skyrmionic systems, where $\tau_0$ assumes unusual values~\cite{rohart_2016,hagemeister_2015}, depends strongly on the mechanism of skyrmion collapse~\cite{bessarab_2018,desplat_2018} and demonstrates extreme sensitivity to applied magnetic field~\cite{bessarab_2018,malottki_2019,wild_2017}. 
Clearly, assuming $\tau_0$ to have some fixed value, e.g. associated with the Larmor precession, can lead to incorrect conclusions about thermal stability of magnetic skyrmions. Recent studies 
have revealed an important role of  
the skyrmion's internal modes in the unusual behaviour of the prefactor~\cite{malottki_2019,desplat_2018}. 
In particular, the modes can cause large entropy barriers, lowered attempt frequencies and enhanced thermal stability of skyrmions~\cite{malottki_2019,desplat_2018,wild_2017,hagemeister_2015}. While knowledge about the Arrhenius prefactor in skyrmionic systems is of great importance for both basic science and technological applications, the relation of the prefactor to the materials fundamentals and the underlying physics have virtually been unknown so far. 

In this article, it is explored to what extent the lifetime of nanoscale, isolated skyrmions in ultrathin ferromagnetic films can be enhanced 
under ambient conditions. 
Fixing the skyrmion size still leaves a space for the optimization of the skyrmion stability by tuning the skyrmion shape, and this possibility is systematically analyzed using the atomistic spin Hamiltonian and harmonic transition state theory~\cite{bessarab_2012,bessarab_2013}, where the curvature of the configuration space arising due to constraints on the length of the magnetic moments is conveniently taken into account by use of general tangent space coordinates and projection operator approach. 

In contrast to previous studies, the analysis of the skyrmion stability diagram goes beyond the evaluation of the collapse energy barrier $\Delta E$ and involves definite calculations of the pre-exponential factor $\tau_0$ instead of treating it as a phenomenological parameter. An extreme sensitivity of the prefactor to magnetic interactions is discovered and explained, thus providing a deep insight into the skyrmion stabilization. Thanks to the pronounced material dependence of the prefactor, it is actually possible to realize long-lived sub-10 nm skyrmions in ferromagnetic films at room temperature and zero applied magnetic field. This finding contrasts sharply with conclusions of previous studies where the skyrmion stability is assessed exclusively based on estimation of the energy barrier. 
Although it is indeed unfeasible to reach energy barriers exceeding thermal energy by a factor of 40-50 at room temperature -- a commonly used criterion for reliable information storage -- while keeping the skyrmion size at nanoscale, the long lifetime of ultrasmall skyrmions can still be achieved due to the remarkably large value of the Arrhenius prefactor $\tau_0$, which is a unique phenomenon in magnetism. This stabilization scenario is particularly realized for skyrmions with a bubble-like profile providing a large number of skyrmion deformation modes and, thereby, high entropy barriers.

\section*{Results}

\noindent
\textbf{Phase diagram.}  
An extended ultrathin skyrmionic system is described here by a classical atomistic spin Hamiltonian on a monolayer hexagonal lattice. 
The total energy $E$ 
of the system 
includes three contributions 
\begin{equation}
    E = E_{\text{ex}} + E_{\text{DM}} + E_{\text{ani}}
    \label{eq:hamiltonian}
\end{equation}
due to the Heisenberg exchange, Dzyaloshinskii-Moriya (DM) interaction and magnetic anisotropy, respectively, 
where each contribution is 
characterized by an effective interaction parameter: $J$, $D$, and $K$ (see Methods section for the detailed description of the simulated system). The technologically-relevant case of zero applied magnetic field is in focus of the present study, therefore the Zeeman term is not included in Eq.~(\ref{eq:hamiltonian}). 
A hexagonal lattice is used here so as to directly mimic ultrathin systems where the presence of skyrmions has been confirmed experimentally~\cite{heinze_2011,romming_2013,romming_2015,hanneken_2015,meyer_2019}, but the results of this work can be interpreted in terms of parameters of the square-lattice model or 
continuous magnetization Hamiltonian using the material parameter transformations provided in Supplementary Note 1.

\begin{figure}
 \includegraphics[width=\columnwidth]{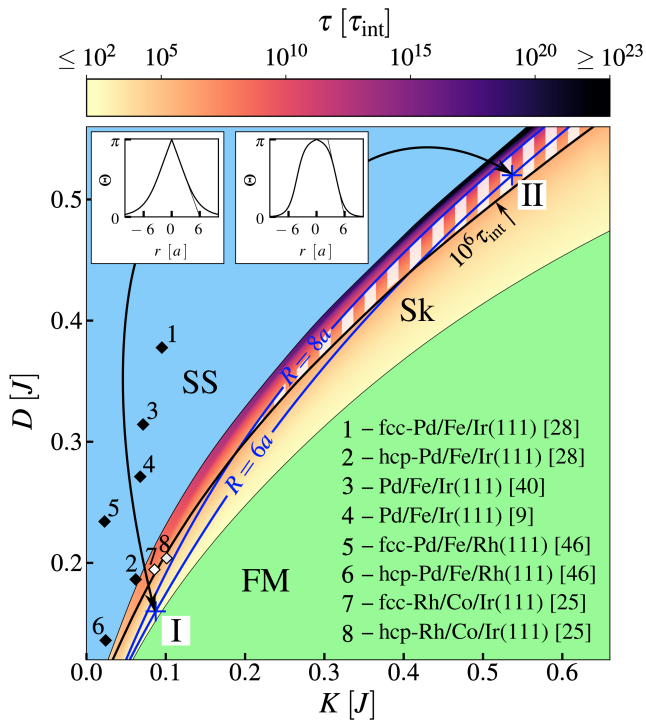}
\caption{\label{fig:diagram} {\bf Magnetic phase diagram for a monolayer of magnetic moments on a hexagonal lattice.} 
A spin spiral is the ground state of the system in the SS region (blue), while the ferromagnetic state is the only stable configuration in the FM domain (green). Metastable, isolated skyrmions in the FM background exist in the Sk sector. 
The color scheme (see top bar) shows the calculated HTST lifetime of isolated skyrmions in terms of the intrinsic precession time $\tau_\text{int} = \mu (J\gamma)^{-1}$, which is on the order of femtoseconds for a typical magnetic system. A thermal energy of $2.6J$ was assumed in the lifetime calculations. Some of the contours of equal skyrmion radius $R$ and lifetime $\tau$ are shown with blue and black solid lines, respectively. Material parameter domain corresponding to skyrmions with $R<8a$ and $\tau>10^6\tau_\text{int}$ is marked with hatching. 
Insets  
show skyrmion profiles and definition of the skyrmion radius for the two points, marked I and II, lying on the same $R$-isoline. Black and white diamonds indicate effective material parameters for 
Fe- and Co-based ultrathin skyrmionic systems, respectively, which are listed in the legend. For Ref.~\cite{meyer_2019}, the parameters were extracted using the method described in Supplementary Note 2.
}
\end{figure}

Figure~1 shows the phase diagram of the system in zero applied magnetic field. 
The diagram was obtained by relaxing a trial single-skyrmion profile to a local energy minimum and examining how the resulting magnetic configuration depends 
on the reduced parameters of magnetic anisotropy, $K/J$, and DM interaction, $D/J$. For the sake of generality, the parameters are varied over a wide range of values, but the physical origin of the parameter variation is beyond the scope of the present study. 
Within a certain domain in the material parameter space, isolated N\'eel-type skyrmions emerge as metastable states in the ferromagnetic (FM) background~\cite{Bogdanov1994}. The skyrmion domain occupies a rather narrow region of the phase diagram~\cite{Bogdanov1989,kiselev_2011,varentsova_2018}. In fact, most of skyrmionic materials known so far exhibit spin-spiral states at zero external magnetic field and a magnetic field needs to be applied for skyrmions to start forming. This scenario is realized, for example, in atomic Pd/Fe bilayers on Ir(111)~\cite{malottki_2017, hagemeister_2015, romming_2015} and on Rh(111)~\cite{haldar_2018}. Based on the values of the effective parameters of magnetic interactions one can place these systems on the phase diagram (see black diamonds labeled $1-6$ in Fig.~1). Nevertheless, zero-field skyrmions have been reported in a few systems, including atomic Rh/Co bilayers on Ir(111)~\cite{meyer_2019}, for which the stacking of the Rh layer affects the material parameters (see white diamonds labeled $7$ and $8$ in Fig.~1, which correspond to hcp and fcc stacking, respectively).

Minimum energy path (MEP) calculations (see Methods section) revealed the mechanism of skyrmion annihilation, which turned out to be the same for all material parameter values within the chosen domain. The mechanism corresponds to usual radial collapse, where the skyrmion symmetrically shrinks and eventually disappears~\cite{bessarab_2015,lobanov_2016,malottki_2017,varentsova_2018}. 

Based on the calculated MEPs for skyrmion annihilation, the lifetime $\tau$ characterizing the stability of the skyrmions against thermally activated decay into the FM state was evaluated using Eq.~(\ref{eq:arrhenius}) where both the energy barrier and the pre-exponential factor were calculated according to the harmonic transition state theory (HTST) (see Methods section). 
The thermal energy of $2.6J$ was assumed in the lifetime calculations, which roughly corresponds to room temperature for $J=10$~meV, a typical value of the Heisenberg exchange parameter in ultrathin magnetic films~\cite{romming_2015}. The calculated lifetime as a function of material parameters is superimposed on the phase diagram in the form of a contour plot. For the sake of generality, the lifetime is presented in Fig.~1 in units of the intrinsic precession time $\tau_\text{int} = \mu(J\gamma)^{-1}$, with $\mu$, $\gamma$ being the on-site magnetic moment and gyromagnetic ratio, respectively. $\tau_\text{int}$ lies in the femtosecond range for usual magnets. 

Calculated distribution of the lifetime demonstrates that not all skyrmions are equally stable, and only the skyrmions whose lifetime is larger than the characteristic laboratory time scale can be experimentally detected. This results in effective narrowing of the skyrmion stability region with temperature. The narrowing can be ascribed to a temperature-dependence of the magnetic interactions, but an explicit account for the thermal activation by means of the rate theory provides a deeper insight into this phenomenon. Thermal activation is in fact an important effect that lies behind the dependence of various magnetic characteristics, e.g. the coercivity~\cite{sharrock_1994,moskalenko_2016}, on temperature. 

The radius $R$ of relaxed, energy-minimum skyrmion states was also quantified using the definition of Bogdanov and Hubert~\cite{bogdanov_1994}:
\begin{equation}
R = r_0-\Theta_0\left[\frac{d\Theta(r)}{d r}\right]_0^{-1},
\label{eq:size}
\end{equation}
where $\Theta(r)$ is the polar angle of the spin located at the distance $r$ from the skyrmion center and the subscript $0$ denotes the point of the steepest slope of $\Theta(r)$. Eight contours of constant $R$ were carefully traced, with $R$ ranging from $5a$ to $12a$ (see Methods Section), $a$ being the nearest-neighbor distance. All the contours correspond to metastable skyrmions with respect to the FM ground state.  
Two of the contours are shown in Fig.~1 as examples.

\begin{figure*}[!ht]
 \includegraphics[width=\textwidth]{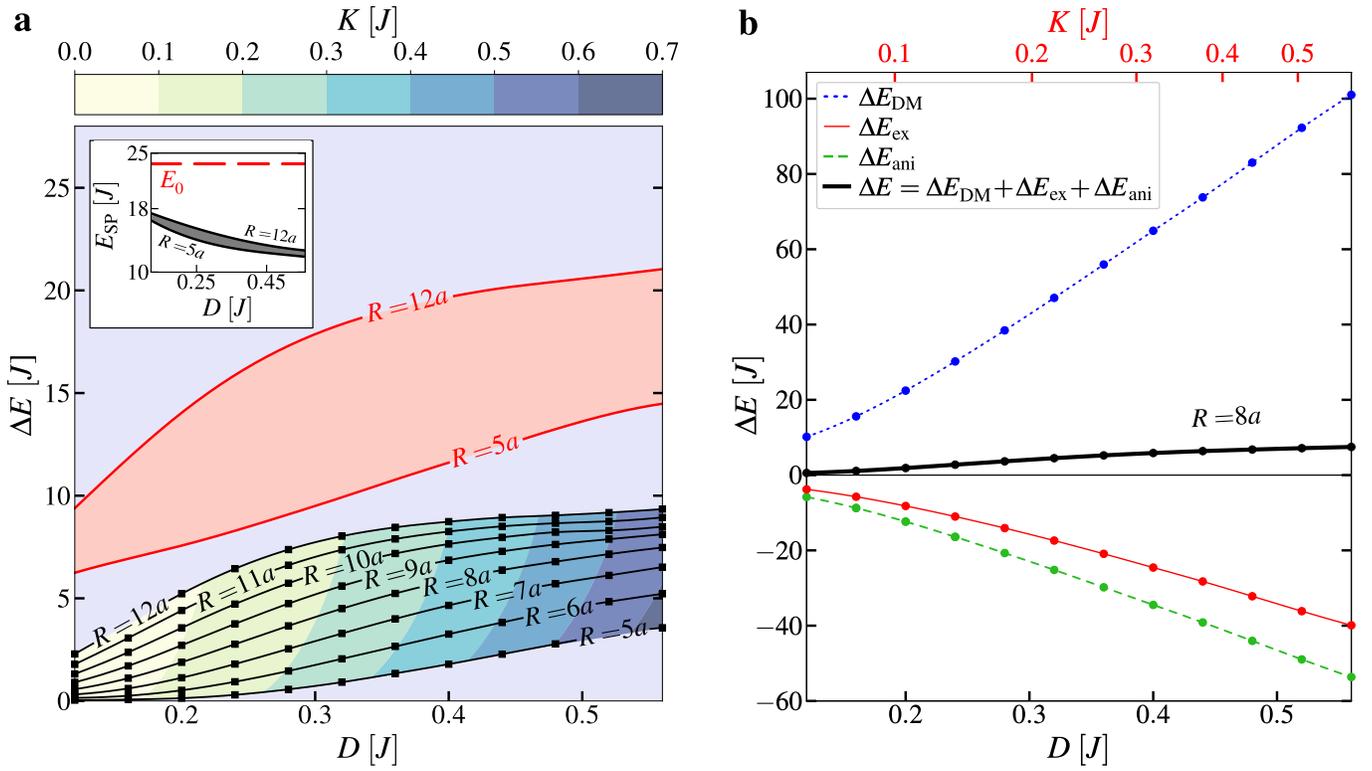}
\caption{\label{fig:barrier} {\bf Results of energy barrier calculations.} {\bf a,} Variation of the GNEB-calculated energy barrier $\Delta E$ for the skyrmion collapse into the FM state (black solid lines) along several contours of equal skyrmion radius $R$. Color codes distribution of the reduced anisotropy parameter. Filled squares indicate the calculated data points. The magnitude of the energy difference between the Belavin-Polyakov soliton state and relaxed, energy-minimum skyrmion state is within the pink area for all skyrmion radii from $5a$ to $12a$. The inset shows the variation of the saddle point energy (dark grey area) and universal energy $E_0$ of the Belavin-Polyakov soliton (red dashed line) relative to the FM state along the $R$-isolines. {\bf b,} Variation of interaction-resolved contributions to the collapse energy barrier due to magnetic exchange $\Delta E_{\text{ex}}$, DM interaction $\Delta E_{\text{DM}}$ and magnetic anisotropy $\Delta E_{\text{ani}}$ along $R=8a$ isoline. Filled circles indicate the calculated data points.}
\end{figure*}

Distributions of the lifetime and radius can now be compared straightforwardly, giving a clear picture of the relationship between skyrmion size and skyrmion stability. 
Both $R$ and $\tau$ increase with the DM interaction strength but decrease as the anisotropy parameter gets larger (see Fig.~1), i.e. both quantities behave alike as functions of relevant material parameters. 
This is consistent with the experimentally observed trend that large skyrmions tend to be more stable than small ones. A more detailed analysis shows, however, that $R$- and $\tau$-isolines do intersect indicating that skyrmion lifetime at a given temperature is not uniquely defined by the skyrmion size. 
In other words, the stability of 
skyrmions can be tuned by a concerted material parameter transformation that preserves the skyrmion radius. This conclusion is further supported by a direct comparison of lifetimes of the skyrmions that belong to the same $R$-contour. For example, parameter sets $K_\text{I}=0.09J$, $D_\text{I}=0.16J$ and $K_\text{II}=0.54J$, $D_\text{II}=0.52J$ (the sets are marked with crosses on the phase diagram, see Fig.~1) result in the same equilibrium skyrmion radius, $R = 6a$. However, the corresponding HTST-estimates of the lifetime differ by five orders of magnitude: $\tau_\text{I}=4.35\times 10^2\tau_\text{int}$, while $\tau_\text{II}=2.44\times 10^7\tau_\text{int}$. 
Due to crossing of $R$- and $\tau$-isolines, it is in fact possible to indicate a material parameter domain which corresponds to skyrmions with $R<R_\alpha$ and $\tau>\tau_\alpha$ for a given time span $\tau_\alpha$ and radius $R_\alpha$. Such a domain is highlighted in Fig.~1 for $\tau_\alpha=10^6\tau_\text{int}$ and $R_\alpha=8a$ as an example. 
 
Interestingly, the shape of the skyrmion profile is not conserved along the $R$-isolines 
either. 
Specifically, the profile has an arrow-like shape for parameter set I; For set II, the circular domain wall encompassing the skyrmion core becomes thinner and the profile changes to a form resembling a magnetic bubble (see the insets in Fig.~1). For a given skyrmion size, the stability of bubble-like skyrmions is in general enhanced compared to that of arrow-like skyrmions. This observation is valuable for the practical 
realization of long-lived, nanoscale skyrmions at room temperature.

It seems clear from Fig.~1 that the skyrmion lifetime can indeed be enhanced via the concerted increase in both $K/J$ and $D/J$ while keeping the skyrmion size unchanged. To elucidate the origin of the lifetime enhancement, it is informative to examine the collapse energy barrier $\Delta E$ and the Arrhenius pre-exponential factor $\tau_0$ separately as functions of displacement along the contours of equal skyrmion radius $R$. This analysis is presented in the following.

\vspace{\baselineskip}\noindent
\textbf{Collapse energy barriers.} 
The variation of $\Delta E$ along the contours of equal skyrmion radius is shown in Fig.~2.  
For all considered radii, the barrier grows monotonically as $K/J$ and $D/J$ increase in a concerted way, i.e. the shape of the skyrmion gradually changes from arrow to bubble.  
Decomposition of the energy barrier into individual interaction-resolved components (see Fig.~2{\bf b}) demonstrates positive contribution from the DM interaction and negative contribution from the Heisenberg exchange and magnetic anisotropy.  
This result is in agreement with earlier studies of the skyrmion collapse~\cite{lobanov_2016,stosic_2017,varentsova_2018}. Although all of the contributions increase steadily in absolute value along the $R$-isolines and largely compensate each other, the balance between them changes, which results in the barrier enhancement.  

Enhancement of the energy barrier along the $R$-isolines is the consequence of the mechanism of skyrmion annihilation. Indeed, progressive shrinking of the skyrmion is produced by a symmetrical rotation of the spins towards the FM state~\cite{bessarab_2015,lobanov_2016,bessarab_2018}. 
Increase in the amplitude of this rotation as well as in the number of spins involved in the process favor enhancement of the corresponding energy barrier.  
This is consistent with the tendency of larger skyrmions to have a higher energy barrier for collapse. 
On the other hand, the overall rotation of spins involved in the radial collapse can be enlarged without changing the skyrmion size simply by making the skyrmion core thicker. This change in the skyrmion shape, leading to the energy barrier enhancement, is indeed realized along the skyrmion radius isolines studied here. As the size of the skyrmion core can not exceed the skyrmion diameter, the growth of the barrier is expected to stop at a certain level for a given skyrmion radius. Figure~2 shows that the barrier enhancement indeed becomes weaker for large values of $K/J$ and $D/J$ corresponding to bubble-like skyrmions for all considered skyrmion radii. 

Within the HTST, the bottleneck for the skyrmion collapse is represented by the relevant first order saddle point (SP) on the multidimensional energy landscape of the system and the energy barrier $\Delta E$ is defined as
\begin{equation}
\Delta E = E_\text{SP}-E_\text{min}, 
\label{eq:barrier}
\end{equation}
where $E_\text{SP}$ and $E_\text{min}$ are the energies of the SP state and the skyrmion state minimum, respectively. 
MEP calculations have shown that SPs for the radial collapse correspond to a very small, Bloch point-like defect in the FM background~\cite{bessarab_2015,lobanov_2016}. With this knowledge, it seems reasonable 
to approximate the bottleneck state by a zero-size skyrmion, whose energy $E_0$ is given by the universal energy of a Belavin-Polyakov soliton with topological charge $|Q|=1$~\cite{belavin_1975,tretiakov_2007}, and define the barrier as $E_0-E_\text{min}$, as it was suggested in previous studies~\cite{buttner_2018}. This method is, however, not justified by the rate theories, and care must be taken when applying it to the skyrmion stability problem. Indeed, $E_0$, which amounts to $4 \sqrt{3} \pi J$ for the atomistic spin model on a hexagonal lattice (see Supplementary Note 3), is significantly larger than the SP energy obtained from the MEP calculations for all skyrmions considered in this study (see the inset in Fig.~2{\bf a}). As a result, the approximate method, although correctly capturing the trend for the barrier, 
gives a large overestimate as compared to the more accurate HTST predictions (see Fig.~2{\bf a}). It is concluded here that the barrier estimates based on the energy of the zero-diameter skyrmion can be inadequate for predicting the skyrmion stability and lifetime quantitatively. This conclusion is supported by the recent study by Heil {\it et al.}~\cite{heil_2019} demonstrating that the Belavin-Polyakov energy needs to be properly corrected in order to obtain an accurate result for the SP energy.

The increase in the energy barrier along the $R$-isolines 
partially explains why bubble-like skyrmions are more stable than arrow-like ones~\cite{varentsova_2018}. However, it is evident from Fig.~2 that the barrier does not exceed $10J$ even for largest skyrmions considered here ($R=12a$). For $J$ on the order of 10 meV, this translates into roughly $4k_BT$ at ambient conditions, which seems to be too little to ensure room-temperature stability on macroscopic time scale~\cite{buttner_2018}. 
It will be shown below that the stabilizing effect of the 
barrier is indeed insufficient for the small skyrmions considered here to be stable at room temperature. However, a long lifetime of nanoscale skyrmions at ambient temperature can still be achieved thanks to extremely large prefactor $\tau_0$, as demonstrated in the following Sections. 

\begin{figure}
 \includegraphics[width=\columnwidth]{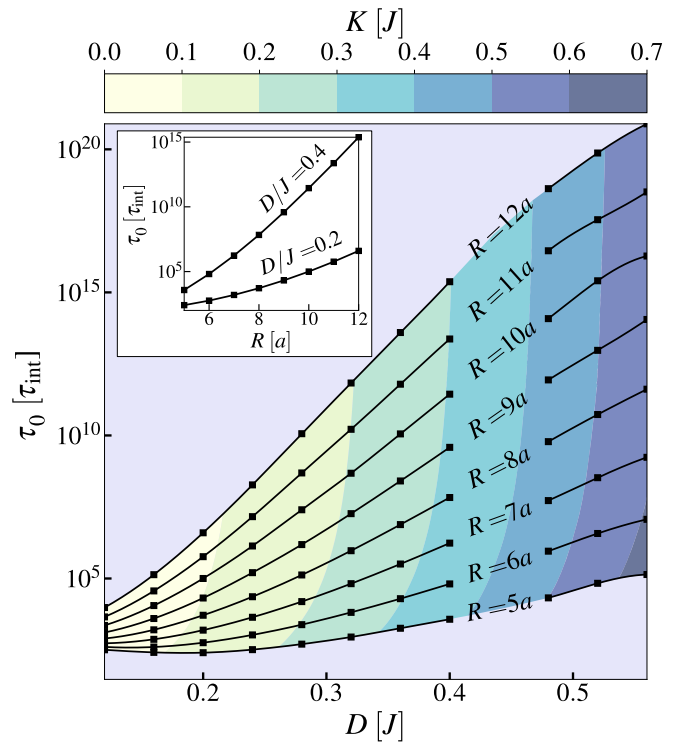}
\caption{\label{fig:prefactor} {\bf Results of pre-exponential factor calculations within HTST.} Variation of the Arrhenius pre-exponential factor $\tau_0$ along several contours of equal skyrmion radius $R$. Color codes distribution of the reduced anisotropy parameter. The inset shows $R$ dependence of $\tau_0$ for the two values of the reduced DM interaction parameter. Filled squares indicate the calculated data points.}
\end{figure}

\begin{figure*}[ht!]
 \includegraphics[width=\textwidth]{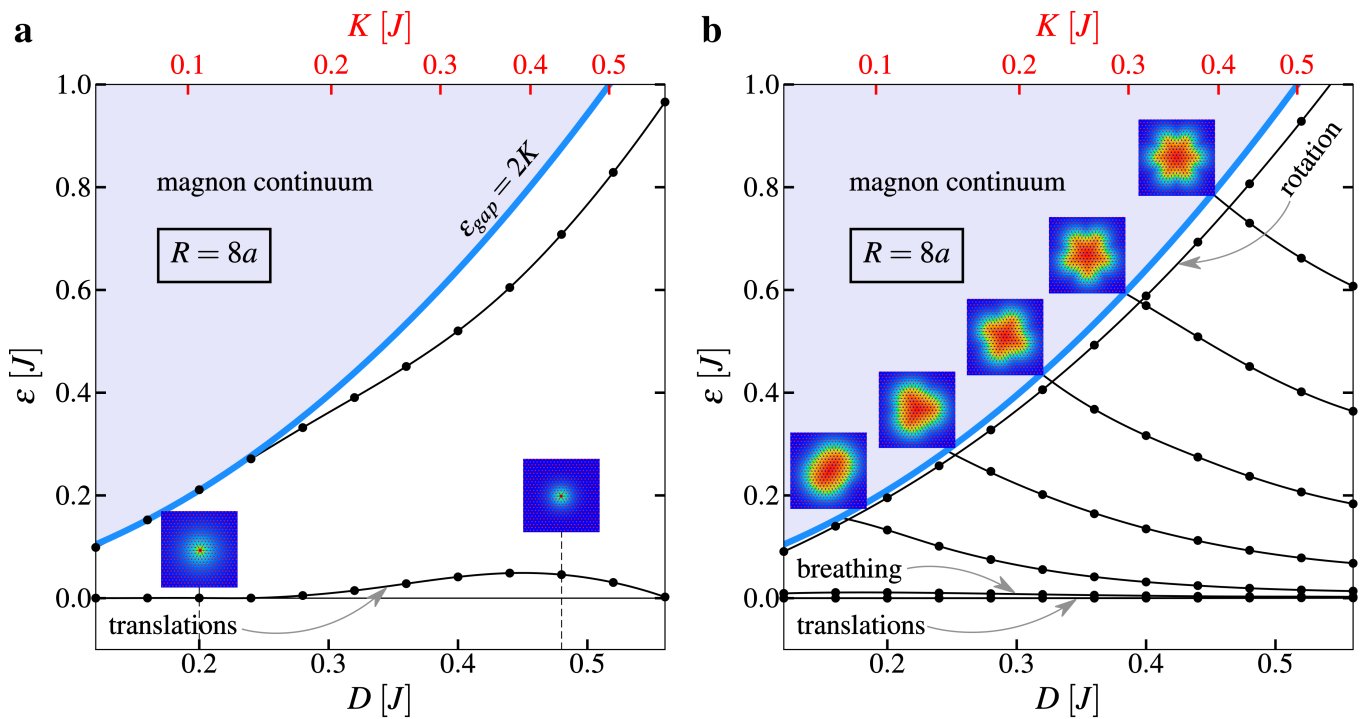}
\caption{\label{fig:spectra} {\bf Variation of the magnetic excitation spectrum at the transition state and at the skyrmion state along one of the contours of equal skyrmion radius.} The spectrum is represented by the eigenvalues of the Hessian matrix calculated at the saddle point ({\bf a}) and skyrmion state minimum ({\bf b}). The calculated data corresponds to skyrmions with $R=8a$. Blue lines show the lower boundary of the continuous part of the spectra, $\varepsilon_{\text{gap}}=2K$. Magnitude of the eigenvalues corresponding to the localized modes are shown with black solid lines. 
Filled circles indicate the calculated data points. 
The insets illustrate the saddle-point configurations ({\bf a}) as well as the skyrmion deformation modes ({\bf b}). The color in the insets indicates the value of the out-of-plane component of the magnetic vectors (red $\leftrightarrow$ up, blue $\leftrightarrow$ down). Negative eigenvalue of the Hessian matrix at the SP is not shown in {\bf a}.}
\end{figure*}

\vspace{\baselineskip}\noindent
\textbf{Pre-exponential factor.} 
Calculated results for the prefactor $\tau_0$ (see Methods section for the details of HTST calculations) are shown in Fig.~3. Overall, $\tau_0$ depends strongly on the material parameters and demonstrates a steady growth along the isolines of skyrmion radius in the direction of increasing $K/J$ and $D/J$. This behavior of the pre-exponential factor in fact enhances the stabilizing effect of the energy barrier for bubble-like skyrmions, i.e. contributions of $\tau_0$ and $\Delta E$ to the skyrmion lifetime are 
both in the same direction. 
Strong variation of $\tau_0$ with the material parameter values confirms that the constant prefactor approximation can indeed be inapplicable for the skyrmion lifetime calculations, which is in agreement with the conclusions of previous studies~\cite{bessarab_2018,malottki_2019,desplat_2018,hagemeister_2015}. Remarkably, the calculated values of $\tau_0$ can span twenty orders of magnitude within the chosen material parameter range. These vast changes in the prefactor are actually comparable to what has recently been observed experimentally for skyrmions in the Fe$_{1-x}$Co$_x$Si system~\cite{wild_2017}.

Inspection of the results obtained for various $R$-isolines also shows sensitivity of $\tau_0$ to the skyrmion radius. 
For fixed $D/J$, the prefactor grows with $R$, contributing to the lifetime enhancement for large skyrmions (see the inset in Fig.~3). 
The prefactor depends on $R$ stronger for larger values of the reduced DM interaction. Strong $R$-dependence of the prefactor was also reported in Ref.~\cite{malottki_2019}, but there the change in the skyrmion size was induced by an external magnetic field rather than by the variation of the material parameters.

The dramatic change in the prefactor stems from the entropic effects~\cite{malottki_2019,desplat_2018,wild_2017,hagemeister_2015}, which are represented within HTST by the Hessian of the energy of the system at the SP and skyrmion state minimum (see Methods section):
\begin{equation}
\label{eq:prefactor_i}
\tau_0 \propto \sqrt{\frac{\det^\prime H_{\text{SP}}}{\det H_{\text{Sk}}}},
\end{equation}
where the determinants can be computed as a product of the eigenvalues. 
The numerical results for the prefactor can be understood by analysing the eigenvalues of the Hessian representing the energy spectra of magnetic excitations at the skyrmion state as well as at the transition state. 

Figure~4 shows how the excitation spectra evolve along one of the $R$-isolines. 
At the transition state, the spectrum only slightly differs from that of the unperturbed FM system due to a rather small 
non-collinear region at the SP 
(see insets in Fig.~4{\bf a}). 
In particular, only a few modes localized on the defect are present within the anisotropy-induced gap, $\varepsilon_{\text{gap}}=2K$, and the number of such modes 
does not change with the material parameter values. 
In contrast, the presence of a skyrmion can produce significant changes to the spectrum of the system~\cite{lin_2014,schutte_2014,psaroudaki_2017,kravchuk_2018}. In addition to the modes corresponding to the in-plane translation and uniform breathing of the skyrmion as well as rotation of the skyrmion core, progressively more localized modes describing various skyrmion deformations emerge in the gap as $K/J$ and $D/J$ increase in a concerted way while the equilibrium skyrmion radius remains unchanged~(see Fig.~4{\bf b}). The deformation modes can be described as periodic modulations of local skyrmion radius along the azimuthal direction. 
For fixed material parameters, the deformation with more modulation periods along the perimeter of the skyrmion corresponds to a larger eigenvalue of the Hessian. On the other hand, the energy of each deformation mode decreases monotonically along the $R$-isoline in the direction of increasing $K/J$ and $D/J$. 
As a result of the difference between the spectra at the transition state and the
skyrmion state, the ratio of the products of eigenvalues -- i.e. the ratio of determinants -- becomes large [see Eqs.~(\ref{eq:prefactor_i}), (\ref{eq:prefactor})]. 
The increase in the number of localized modes in the magnon gap and their softening, which are characteristic features of the skyrmion state, do not occur at the transition state. This increases the entropy difference between the skyrmion state and transition state which 
in turn leads to larger values of the pre-exponential factor $\tau_0$ for the skyrmions with the bubble-like shape. 

\vspace{\baselineskip}\noindent
\textbf{Skyrmion lifetime.} 
Figure~5 summarizes the calculated results for the lifetime $\tau$ of nanoscale isolated skyrmions in an ultrathin FM film at ambient conditions. 
Absolute values for $\tau$ were obtained by setting temperature $T$ to 300~K and assuming typical values for the exchange parameter $J$ and the magnitude of the 
magnetic moment~$\mu$: $J=10$~meV and $\mu=3\mu_\text{B}$, with $\mu_\text{B}$ being Bohr magneton~\cite{romming_2015}. 
Strong variation of the prefactor along the $R$-isolines (see Fig.~3) has a clear impact on the skyrmion lifetime, which can exceed the level of ten years for relatively large reduced DM interaction and anisotropy. Note that the diameter of the skyrmions considered here does not exceed $24a$, which translates into 6.5 nm for the samples grown on the Ir(111) surface~\cite{heinze_2011}. Therefore, it is still possible to obtain long skyrmion lifetime for more moderate values of the material parameters by increasing the skyrmion size while keeping it within the nanoscale range. 

\begin{figure}[ht!]
 \includegraphics[width=\columnwidth]{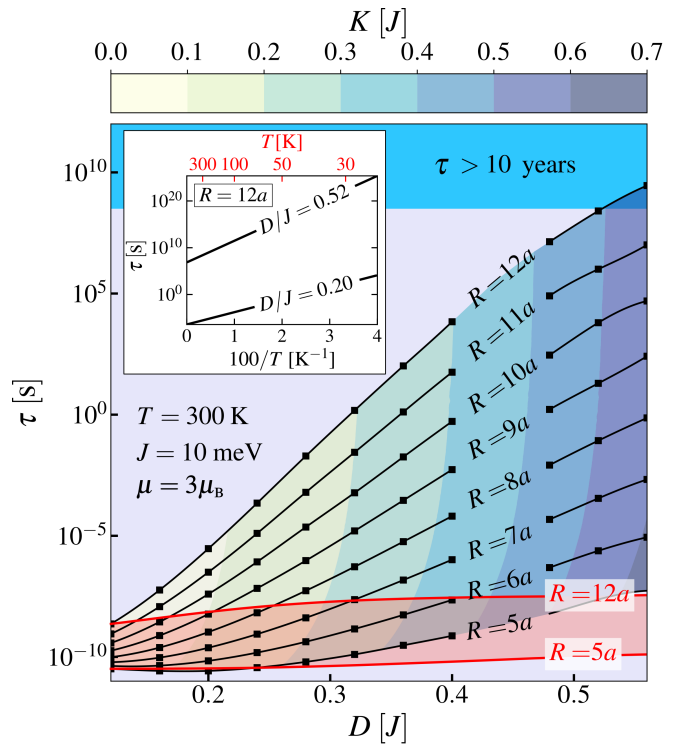}
\caption{\label{fig:lifetime} {\bf Results of skyrmion lifetime calculations.} Variation of the HTST-predicted skyrmion lifetime $\tau$ (black solid lines) along several contours of equal skyrmion radius $R$. The following values of temperature $T$, Heisenberg exchange parameter $J$ and magnitude of the magnetic moments $\mu$ were assumed in the calculations: $T=300$~K, $J=10$~meV, $\mu=3\mu_\text{B}$. Color codes distribution of the reduced anisotropy parameter. All calculated data points indicated by filled squares correspond to metastable skyrmions with respect to the FM ground state. The inset shows the Arrhenius plot for the two values of the reduced DM interaction parameter on $R=12a$ isoline. The magnitude of the lifetime calculated in the constant Arrhenius prefactor approximation is within the pink area for all skyrmion radii from $5a$ to $12a$. Blue area indicates lifetimes greater than ten years.}
\end{figure}

To emphasize the decisive role of the Arrhenius pre-exponential factor in the stabilization of nanoscale skyrmions at room temperature, the lifetime 
was also calculated using the commonly applied constant prefactor approximation. 
For each skyrmion radius, the value of $\tau_0$ calculated at $D/J=0.12$ was used to evaluate the lifetime along the entire $R$-isoline.
For $D/J=0.12$, the magnitude of $\tau_0$ changes from 30~ps for $R=5a$ to 1~ns for $R=12a$, which is actually in the range of typical prefactor values used in the literature to calculate the rate of thermally activated magnetic transitions~\cite{stosic_2017,weller_1999,chen_2010,bedanta_2008,kapaklis_2014}. The values of the skyrmion lifetime calculated with these values of $\tau_0$ are enclosed in the pink area in Fig.~5. Since the prefactor is assumed to remain constant along each $R$-isoline in this case, the slight increase in the lifetime originates from the increase in the energy barrier. Clearly, this lifetime enhancement is insignificant compared to what HTST predicts. 
Relatively small increase in the energy barrier but a large change in the pre-exponential factor are further illustrated by the Arrhenius plots for the two points along $R=12a$ isoline (see the inset in Fig.~5): The plots have similar slopes but intercept the vertical axis at very different levels.

\section*{Discussion}

\noindent The qualitative difference in the behavior of the excitation spectra at the transition state and at the skyrmion state is the origin of large changes in the prefactor. This phenomenon is observed for all skyrmion sizes considered in the present study (see Supplementary Fig.~1). However, larger skyrmions can accommodate more modulation periods along their circumference. This leads to stronger variation of $\tau_0$ along $R$-isolines corresponding to larger skyrmion radii (see Fig.~3). The possibility to introduce more localized modes in the magnon gap by increasing the skyrmion size also explains the sensitivity of the prefactor to the skyrmion radius. 

The emergence of the modes localized on the skyrmion and their uncompensated softening are the primary reasons for the large variation of the prefactor along the $R$-isolines. To gain a qualitative insight into these effects, it is convenient to treat an isolated skyrmion as a composition of a core, an outer FM domain, and a 
wall separating the core and the outer domain. Under this representation, transformation of the skyrmion shape along the $R$-isoline can be interpreted in terms of variation of the domain wall (DW) width, with wider DWs corresponding to arrow-like skyrmions and thinner DWs corresponding to bubble-like skyrmions (see the insets in Fig. 1). On the other hand, the skyrmion deformations can be viewed as transverse fluctuations of the DW and parametrized by Fourier harmonics. Neglecting the curvature of the DW and disregarding the DM interaction as well as long-range dipolar interaction, the deformation energy $\varepsilon_n$ associated with the $n$th harmonic is given by~\cite{makhfudz_2012}:
\begin{equation}
\label{eq:trans_mode}
\varepsilon_n = \frac{6\pi^2a^2J}{L^2}n^2,
\end{equation}
where $L$ is the DW length~(see Supplementary Note 4). 
The number of harmonics $N$ in the magnon gap is then defined by the largest $n$ for which $\varepsilon_n<\varepsilon_{\text{gap}}=2K$: 
\begin{equation}
N \propto \frac{L}{\Delta},
\end{equation}
where $\Delta \sim a\sqrt{J/K}$ is the DW width. $N$ gives a qualitative estimate of the number of the modes localized on the skyrmion with perimeter $2\pi R=L$. This increases monotonically as the DW becomes thinner, which explains why progressively more modes localize on the skyrmion as it gradually changes its shape along the $R$-isoline to form a bubble-like structure. 

Equation~(\ref{eq:trans_mode}) does not explain well all features of the excitation spectrum of the skyrmion because it is based on the assumption of straight, achiral DW. 
In particular, the deformation mode softening for the bubble-like skyrmions is not reproduced by Eq.~(\ref{eq:trans_mode}). However, the mode softening can still be obtained using the straight DW model if one includes the DM interaction in the Hamiltonian and considers a concerted increase in both $K$ and $D$. In particular, energy of transverse deformation modes of the DW does show a decrease when the variation of $K$ and $D$ is identical to that along the contours of equal skyrmion radius~(see Supplementary Fig.~2).

It seems clear from the above 
results that room temperature stability of nanoscale skyrmions can be achieved almost entirely due to exceptionally large pre-exponential factor $\tau_0$ rather than high energy barrier $\Delta E$. 
Large values of the prefactor are an indication of large entropy barriers~\cite{wild_2017,hagemeister_2015,desplat_2018,malottki_2019}. Therefore, the concept of obtaining long skyrmion lifetime by means of the enhanced entropy barriers, which was pointed out independently in Refs.~\cite{desplat_2018,malottki_2019}, becomes critical for sub-10 nm skyrmions at room temperature. 
A key strategy for the realization of large entropy barriers stabilizing the skyrmion state is to establish as many localized modes corresponding to skyrmion deformations as possible, because these modes introduce uncompensated increase in the entropy of the skyrmion state. For nanoscale skyrmions at zero applied magnetic field, the sufficient number of localized modes is realized when the skyrmions resemble magnetic bubbles, i.e. when 
reduced 
parameters of DM interaction and anisotropy acquire relatively large values compared to the Heisenberg exchange parameter. 
This regime seems difficult to achieve in typical transition-metal systems where $J$ assumes usual values on the order of 10~meV. However, desired enhancement of $D/J$ and $K/J$ can actually be obtained 
by reducing the strength of the Heisenberg exchange (calculated skyrmion lifetime for several values of $J$ and $T$ is shown in Supplementary Fig.~3). For example, rather large values of $D/J = 1.3$ and $K/J = 0.3$ have been achieved in the [RhPd/2Fe/2Ir] system thanks to relatively low value of $J = 2$~meV~\cite{dupe_2016}. 
Note that the parameter $J$ should be understood as an effective exchange constant for systems with frustrated exchange. There, the effective exchange parameter characterizes well low-energy excitations of the system, but may fail to describe the states far from the energy minimum~\cite{malottki_2017}, and, therefore, small value of $J$ does not necessarily mean low ordering temperature~\cite{bottcher_2018}, especially for the anisotropic systems~\cite{bruno_1991}. 

Interestingly, the material parameters approach the desired values in ultrathin Fe and Co films (see Fig.~1) which have already
been studied~\cite{malottki_2017,hagemeister_2015,romming_2015,haldar_2018,meyer_2019}, making these systems promising for ultrasmall, room-temperature
skyrmions stabilized by entropy barriers. 
Tuning of the magnetic interactions can be achieved via various mechanisms. For example, the DM interaction can be engineered at interfaces between ferromagnet and heavy metal~\cite{yang_2015}, or graphene~\cite{yang_2018} and controlled by 3{\it d}-band filling in weakly ferromagnetic insulators~\cite{beutier_2017}. Adjustment of the magnetic anisotropy or/and Heisenberg exchange is also important for the enhancement of the skyrmion stability without changing the skyrmion size. 
Thin films of 3{\it d} elements sandwiched between 4{\it d} and 5{\it d} transition metals appear to be particularly interesting systems since they make it possible to tune $J$, $D$ and $K$ independently and over a wide range via layer composition, alloying, and
intermixing at the 4{\it d}/3{\it d} and 5{\it d}/3{\it d} interfaces. The feasibility of such local tuning of magnetic interactions by hybridization at 3{\it d}/4{\it d} and 3{\it d}/5{\it d} interfaces has been demonstrated based on density functional theory calculations for [Rh$_x$Pd$_{1-x}$/Fe/Ir] 
structures~\cite{dupe_2016} reporting an order of magnitude variation of the effective exchange interaction parameter and large changes in the magnetocrystalline anisotropy in the magnetic layer. 
Large variations of the DM interaction have been reported for [Rh/Co/Pt] systems~\cite{jia_2018}. Interface tuning of the magnetic interactions is a local effect based on the hybridization which does not rely on the multilayer structure. Therefore, it is relevant for ultrathin film systems, too. Possibility to tune exchange interactions has indeed been demonstrated for Co-based ultrathin films~\cite{meyer_2019}. It is anticipated that a variation of the composition of $4d/3d/5d$ film structures -- which are yet to be explored experimentally -- could make it possible to reach the regime of isolated zero-field skyrmions which are stable at room temperature.

In conclusion, the present study explored thermal stability of skyrmions in ultrathin ferromagnetic films by means of harmonic transition state theory and atomistic spin Hamiltonian. The study predicted that the small skyrmions can still be quite stable at ambient conditions if the circular domain wall encompassing the skyrmion core becomes thin and the skyrmion profile resembles a magnetic bubble. The bubble-like profile ensures a large number of skyrmion deformation modes in the magnon gap thereby creating a high entropy barrier for the skyrmion collapse and establishing a long skyrmion lifetime due to the large Arrhenius pre-exponential. The regime of high entropy barriers for the nanoscale skyrmions can be achieved by increasing the magnetic anisotropy concertedly with the DM interaction or in fact by decreasing the Heisenberg exchange interaction to reach large values of $K/J$ and $D/J$. The findings of the present study deepen the understanding of skyrmion stability and its relationship with the skyrmion size and provide an avenue for the realization of nanoscale, room-temperature stable skyrmions.

The results presented here will re-establish interest in ultrathin films as technologically-relevant skyrmionic systems. In skyrmionics, the focus has largely been shifted toward multilayered systems comprising heavy-metal/ferromagnet layers, with a large number of repetitions. However, increase in the number of layers in such systems results in large stray fields which can induce twisted spin textures with a nonuniform chirality across the film thickness and, as a result, complicate control over the magnetic structure. Overall increase in the thickness of the system also increases power dissipation associated with current-driven skyrmion motion~\cite{juge_2019}. These unfavorable effects can be avoided in ultrathin film systems, where nanoscale skyrmions at ambient conditions can be stabilized thanks to ultralow attempt frequency.

The present study provides a general understanding of thermal stability of skyrmions in ferromagnetic films and reveals dependencies on basic parameters relevant for most skyrmionic systems. However, additional terms in the Hamiltonian can affect the skyrmion lifetime. For example, frustration in the magnetic pairwise interactions~\cite{malottki_2017} as well as higher-order exchange~\cite{paul_2020} can enhance the energy barrier protecting the skyrmion from collapsing. The details of the exchange interactions beyond nearest neighbors can influence the Arrhenius prefactor~\cite{bessarab_2018,hoffmann_2020}. Engineering the exchange frustration and higher-order interactions is an additional source of the skyrmion stability enhancement~\cite{malottki_2017,paul_2020,bessarab_2018,hoffmann_2020}. Nevertheless, the principles of skyrmion stabilization derived here using the basic Hamiltonian describing chiral skyrmions in zero applied magnetic field, e.g. the lifetime enhancement by the realization of a large difference between the excitation spectra at the transition state and at the skyrmion state, are rather general and can also be applied to systems characterized by additional magnetic interactions. 

\section*{Methods}

\noindent
\textbf{Simulated system.}
A two-dimensional skyrmionic system is modeled as a single monolayer of classical magnetic vectors localized on vertices of a hexagonal lattice. 
The energy of the system is given by Eq.~(\ref{eq:hamiltonian}) where each contribution is defined as follows:

\begin{align}
   E_{\text{ex}} &= -\frac{J}{2}\sum_{\langle i,j \rangle}\vec{m}_i\cdot \vec{m}_j, \label{eq:ex}\\
   E_{\text{DM}} &= -\frac{D}{2}\sum_{\langle i,j \rangle}\vec{d}_{ij}\cdot\left[\vec{m}_i\times \vec{m}_j\right], \label{eq:dm}\\
   E_{\text{ani}} &= -K\sum_{i} (\vec{m}_i\cdot \vec{e}_K)^2. \label{eq:ani}
\end{align}
Here, 
$\vec{m}_i$ is the unit vector in the direction of the magnetic moment at lattice site $i$; $J$ and $D$ are the parameters of Heisenberg exchange and DM interaction between nearest neighbor spins, respectively. 
The unit DM vector $\vec{d}_{ij}$ lies in the monolayer plane and points perpendicular to the bond connecting sites $i$ and $j$. The model also includes out-of-plane anisotropy with the easy axis defined by the unit vector $\vec{e}_K$ pointing perpendicular to the monolayer plane and effective parameter $K$ incorporating both magnetocrystalline and magnetostatic contributions~\cite{draaisma_1988}. Effective treatment of the stray fields in the context of skyrmion stability is justified for thin films~\cite{lobanov_2016}, but may fail as the film thickness increases.

Only one single skyrmion is placed in the simulated system. The size of the computational domain is chosen to be 80$\times$80 lattice sites, which is large enough for the isolated equilibrium skyrmion solution not to be affected by the boundaries. Periodic boundary conditions are applied so as to model extended two-dimensional systems.

\vspace{\baselineskip}\noindent
\textbf{Identification of isolines of skyrmion radius.}
Each isoline $\alpha$ along which the skyrmion radius $R$~[see Eq.~(\ref{eq:size})] assumes a constant value of $R_\alpha$ was obtained numerically using the following technique. At first, profiles of $R$ as functions of the magnetic anisotropy $K$ for several fixed values of the DM interaction, $D=D_l$, are calculated. For each profile, the sought-for value of the anisotropy parameter $K_l$ corresponding to the predefined skyrmion radius $R_\alpha$ is isolated by interpolation between the data points and then refined using the bisection method until the desired accuracy has been achieved. As a result, a set of 12 points $(K_l,D_l)_\alpha$, $l=1,\ldots,12$, is obtained, where each point corresponds to a skyrmion with the radius $R=R_\alpha\pm 0.1a$. This set of points in the material parameter space gives a discrete representation of the $R$-isoline. Eight isolines corresponding to $R=5a,6a,\ldots,12a$ have been calculated in this manner. They are presented in Supplementary Table 1. 

\vspace{\baselineskip}\noindent
\textbf{Harmonic transition state theory.}
The mean skyrmion lifetime which characterizes the stability of the skyrmion state with respect to thermal fluctuations is calculated using harmonic transition state theory (HTST) for magnetic systems~\cite{bessarab_2012,bessarab_2013} extended to include the presence of Goldstone modes~\cite{bessarab_2018}. The curvature of the configuration space of a magnetic system arising due to the constraints on the length of the magnetic moments is taken into account by use of general tangent space coordinates and projection operator approach, as described in the following. This formulation results in more convenient calculations compared to approaches based on spherical coordinates~\cite{bessarab_2012}. HTST presents a rigorous foundation for the Arrhenius law [see Eq.~(\ref{eq:arrhenius})] and provides means for 
definite 
evaluation of both the energy barrier $\Delta E$ and the pre-exponential factor $\tau_0$ for thermally activated decay processes~\cite{vineyard_1957,bessarab_2012}. Within HTST, analysis of skyrmion stability relies on the identification of relevant saddle points (SPs) on the energy surface~[see Eqs.~(\ref{eq:hamiltonian}),~(\ref{eq:ex})-(\ref{eq:ani})] characterizing transition states for the skyrmion decay into the ferromagnetic state or other states available in the system, as described in the following.

Within the classical transition state theory, the inverse of the lifetime i.e., the rate of thermally activated escape from a metastable state such as the skyrmion state, is given by the following equation:
\begin{equation}
\label{eq:tst}
    \tau^{-1}=\left<\delta[f(\mathbf{m})]v_\perp(\mathbf{m})h[v_\perp(\mathbf{m})]\right>,
\end{equation}
where vector $\mathbf{m}$ defines magnetic configuration of the system, angular brackets denote the thermal averaging with a Boltzmann distribution, $f(\mathbf{m})=0$ defines the transition state dividing surface separating the initial state from the rest of the configuration space, $v_\perp(\mathbf{m})=\mathbf{\nabla}f(\mathbf{m})\cdot\dot{\mathbf{m}}$ is the component of the velocity along the local normal of the dividing surface. Heaviside step function $h[v_\perp(\mathbf{m})]$ signifies that all trajectories pointing away from the initial state at the dividing surface contribute to the escape rate, which is the central assumption of the transition state theory. 

Harmonic approximation to the transition state theory corresponds to taking the dividing surface to be a hyperplane that goes through a first order SP on the energy ridge surrounding the initial state. The hyperplane is oriented in such a way that its normal coincides with the Hessian's eigenvector along which the SP is a maximum. HTST estimate of the mean lifetime is obtained from Eq.~(\ref{eq:tst}) by introducing quadratic approximations to the energy surface around the initial state minimum and the SP and linearizing the Landau-Lifshitz equation to compute $\dot{\mathbf{m}}$ in the neighbourhood of the SP. In particular, after some algebra (which will be published elsewhere) one obtains the Arrhenius expression for the skyrmion lifetime $\tau$ [see Eq.~(\ref{eq:arrhenius})], where the energy barrier $\Delta E$ is given by the SP energy relative to the skyrmion energy minimum [see Eq.~(\ref{eq:barrier})] and the pre-exponential factor is described by the following formula~\cite{bessarab_2012,bessarab_2018}: 

\begin{equation}
\tau_0^{-1}=\frac{\lambda}{2\pi}\frac{V_{\text{SP}}}{V_{\text{Sk}}}\sqrt{\frac{\det H_{\text{Sk}}}{\det^\prime H_{\text{SP}}}}.
\label{eq:prefactor}
\end{equation}
Here, $\lambda$ is the dynamical factor describing the flux of trajectories through 
the surface separating the skyrmion state from the FM state in the configuration space, $\det H_{\text{Sk}}$ and $\det H_{\text{SP}}$ denote the determinants of the Hessian matrices at the skyrmion state minimum and the SP, respectively. The determinants are computed as a product of the eigenvalues and the prime means that the negative one is omitted. At elevated temperatures which are of interest here, in-plane translations of both the skyrmion and SP configuration should be treated as Goldstone modes with volumes $V_{\text{Sk}}$ and $V_{\text{SP}}$, respectively, and corresponding eigenvalues not included in the determinants. The calculation of individual terms involved in the expression for the pre-exponential factor is described in the following. Calculated values of the energy barrier, dynamical factor, ratio of the determinants, Goldstone mode volumes for the skyrmions corresponding to one of the isolines of skyrmion radius are presented in Supplementary Table 2.

Being a quasi-equilibrium theory, transition state theory assumes that Boltzmann distribution is established in the region of configuration space that corresponds to the skyrmion state before the system escapes due to thermal fluctuations. 
This assumption is usually justified when 
the energy barriers are large compared to the thermal energy. A more general prerequisite is that the escape events are rare on the intrinsic time scale of magnetization dynamics of the system. For long-lived skyrmions which are of interest here, this condition is expected to be met {\it a priori}. Note that 
the separation of time scales can arise from both an energy barrier and an entropy barrier. For skyrmions, stabilization by entropy barriers 
is particularly important.

\vspace{\baselineskip}\noindent
\textbf{Minimum energy path calculations.}
Transition state saddle points needed for the skyrmion lifetime calculations are identified by locating a maximum along minimum energy paths (MEPs) for skyrmion collapse. Similar to previous studies (see e.g. Ref.~\cite{varentsova_2018}), the MEPs are calculated using a geodesic nudged elastic band (GNEB) method~\cite{bessarab_2015,bessarab_2017}, an extension of the nudged elastic band method~\cite{henkelman_1998} to magnetic systems. At first, a chain of ten to twenty copies, or images, of the system is created along the shortest path connecting the skyrmion state and the FM state. The equidistant image distribution is achieved by adding virtual springs between adjacent images. The chain is then iteratively brought to the lowest position on the energy surface. The final, relaxed distribution of the images gives a discrete representation of the MEP. The location of the maximum is accurately refined using the climbing image technique~\cite{bessarab_2015,henkelman_2000}. The calculated MEPs for collapse of the skyrmions corresponding to one of the $R$-contours are shown in Fig.~6.

\begin{figure}[ht!]
 \includegraphics[width=\columnwidth]{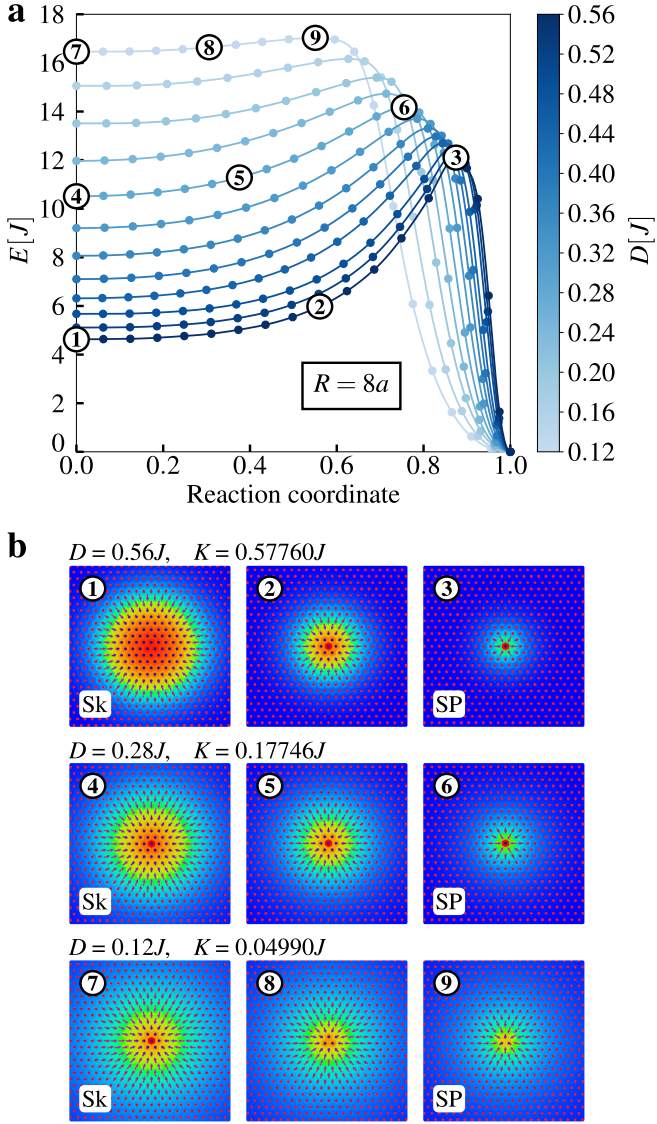}
\caption{\label{fig:mep} {\bf Minimum energy paths for the skyrmion collapse.}  {\bf a,}  Energy variation along the MEPs for radial collapse of the skyrmions that belong to the same contour of equal skyrmion radius, with $R=8a$. The color of the curves codes the magnitude of the DM interaction parameter. The filled circles show position of the intermediate states along the collapse paths. The reaction coordinate is defined as the normalized displacement along the MEP. The starting- and end-points of the reaction coordinate are the skyrmion (Sk) and ferromagnetic states, respectively. The encircled numbers label the states for which spin configurations are shown in {\bf b}. The background color indicates the value of the out-of-plane component of the magnetic vectors (red $\leftrightarrow$ up, blue $\leftrightarrow$ down).
}
\end{figure}

\vspace{\baselineskip}\noindent
\textbf{Evaluation of the Hessian.}
The configuration space of a system of $P$ magnetic moments with fixed length is a 2$P$-dimensional Riemannian manifold, $\mathcal{R}$, represented by a direct product of $P$ 2-dimensional spheres associated with each magnetic moment vector. 
The special form of the configuration space $\mathcal{R}$ must be accounted for in the evaluation of the Hessian. One option is to 
use spherical coordinates, but this approach may suffer from singularities at the poles. Instead, we obtain the Hessian using a projection operator approach~\cite{mueller_2018}. First, the Hessian in the $3P$-dimensional embedding Euclidean space is defined by calculating the matrix of second-order partial derivatives of $E$ with respect to the Cartesian components of the magnetic vectors:

\begin{equation}
\label{eq:hessian_3P}
\mathcal{H}=
\begin{bmatrix}[2.5]
 \dfrac{\partial^2\: E}{\partial m^x_1 \partial m^x_1} & \dfrac{\partial^2\: E}{\partial m^x_1 \partial m^y_1} &  \dots  & \dfrac{\partial^2\: E}{\partial m^x_1 \partial m^z_P} \\
 \dfrac{\partial^2\: E}{\partial m^y_1 \partial m^x_1} & \dfrac{\partial^2\: E}{\partial m^y_1 \partial m^y_1} & \dots  & \dfrac{\partial^2\: E}{\partial m^y_1 \partial m^z_P} \\
 \vdots   & \vdots     & \ddots &\vdots \\
 \dfrac{\partial^2\: E}{\partial m^z_P \partial m^x_1} & \dfrac{\partial^2\: E}{\partial m^z_P \partial m^y_1} & \dots  & \dfrac{\partial^2\: E}{\partial m^z_P \partial m^z_P}
\end{bmatrix}.
\end{equation} 
Note that $\mathcal{H}$ is independent of the magnetic configuration for quadratic Hamiltonians such as the one defined by Eqs.~(\ref{eq:hamiltonian}), (\ref{eq:ex})-(\ref{eq:ani}). 
The sought-for Hessian $H$ in the configuration space $\mathcal{R}$ is obtained using the following equation:
\begin{equation}
    \label{eq:hessian_2P}
    H = U^T(\mathcal{H}-\mathcal{L}) U.
\end{equation}
Here, $U$ is a $3P\times 2P$ matrix projecting onto the local tangent space of $\mathcal{R}$ and $\mathcal{L}$ is a matrix representation of the shape operator~\cite{mueller_2018} accounting for the curvature of the configuration space. For each individual magnetic moment $i$, the shape operator is defined as:
\begin{equation}
    \label{eq:shape_i}
    \mathcal{L}_i = (\vec{m}_i\cdot \vec{\nabla}_i E)I,
\end{equation}
where $\vec{\nabla}_i\equiv \partial/\partial\vec{m}_i$ and $I$ is the $3\times 3$ unit matrix. The shape operator for the whole system is simply a direct sum of all $\mathcal{L}_i$:
\begin{equation}
\label{eq:shape}
\mathcal{L}=\bigoplus_{i=1}^P \mathcal{L}_i\equiv
\begin{bmatrix}
 \mathcal{L}_1 & 0 & \dots  & 0 \\
 0 & \mathcal{L}_2 & \dots  & 0 \\
 \vdots & \vdots & \ddots &\vdots \\
 0 & 0 & \dots  & \mathcal{L}_P
\end{bmatrix}.
\end{equation}
The projection matrix $U$ can be obtained by computing a direct sum of all $3\times 2$ projection matrices $U_i$ associated with each magnetic moment $i$: $U=\bigoplus\limits_{i=1}^P U_i$. Columns of $U_i$ are orthonormal vectors, $\vec{\eta}_i$ and $\vec{\xi}_i$, defining a basis in the tangent space for the magnetic moment $i$. The choice of the basis is arbitrary. 
For example, $\vec{\eta}_i$ can be obtained by orthonormalization of a random vector with respect to $\vec{m}_i$ and then $\vec{\xi}_i$ can be generated using the cross-product: $\vec{\xi}_i=[\vec{\eta}_i\times \vec{m}_i]$. Note that both the shape operator $\mathcal{L}$ and the projector $U$ depend on the magnetic configuration $\mathbf{m}=(\vec{m}_1,\ldots,\vec{m}_P)$, which is in contrast to the Hessian $\mathcal{H}$ in the embedding Euclidean space. 

The projection operator approach was applied to calculate Hessians at the SP and at the skyrmion state minimum, and then the eigenvalues needed to compute the determinants in Eq.~(\ref{eq:prefactor}) were calculated using the Intel Math Kernel Library~\cite{mkl}. 

\vspace{\baselineskip}\noindent
\textbf{Evaluation of the dynamical factor.} The dynamical factor $\lambda$ is computed using the following equation:
\begin{equation}
    \label{eq:lambda}
    \lambda = \frac{\gamma}{\mu}\sqrt{\mathbf{s}^T A^T H_{\text{SP}}\,A\,\mathbf{s}}.
\end{equation}
Here, $H_{\text{SP}}$ is the $2P\times 2P$ Hessian matrix computed at the SP according to Eq.~(\ref{eq:hessian_2P}), $\mathbf{s}$ is the eigenvector of $H_{\text{SP}}$ corresponding to the negative eigenvalue, and $A$ is a $2P\times2P$ block-diagonal matrix computed as a direct sum of the Pauli matrices: $A=\bigoplus\limits_{i=1}^P -\mathrm{i}\sigma_y$, with $\sigma_y$ defined by:

\begin{equation}
    - \mathrm{i}\sigma_y = 
\begin{bmatrix}
0 & -1 \\
1 & 0
\end{bmatrix}.
\end{equation}
Note that the expression for $\lambda$ can also be written in a different form, see Ref.~\cite{potkina_2020} for details.

Expression~(\ref{eq:lambda}) for $\lambda$ in fact does not require calculation of the eigenvectors of $H_{\text{SP}}$. Indeed, the eigenvector $\mathbf{s}$ coincides with the unit tangent to the MEP at the SP, which is available from the GNEB calculations~\cite{bessarab_2015}. Since the tangent vector $\mathbf{t}$ is usually defined in the $3P$-dimensional embedding Euclidean space, its evaluation in the $2P$-dimensional $U$-basis associated with the SP configuration requires the following transformation:
\begin{equation}
    \mathbf{s}=U^T\mathbf{t}.
\end{equation}

\vspace{\baselineskip}\noindent
\textbf{Evaluation of the Goldstone mode volumes.}
The Goldstone modes correspond to translations of localized magnetic structures, i.e. the skyrmion state and the SP state, in the film plane. Consequently, the volumes of the modes can be obtained by integration over spatial coordinate~\cite{braun_1994}, as described below. 
Let $\mathbf{m}_\beta(\vec{r})$ be the magnetic configuration of the skyrmion state ($\beta = \text{Sk}$) or the SP state ($\beta = \text{SP}$) localized at position $\vec{r}$. 
Translation of the magnetic texture along direction $\vec{e}$ by a distance $dr$ can be described by 
\begin{equation}
\mathbf{m}_\beta(\vec{r}+\vec{e}dr)-\mathbf{m}_\beta(\vec{r})=(\vec{e}\cdot \vec{\nabla}_r)\mathbf{m}_\beta(\vec{r})dr.
\label{eq:s1}
\end{equation}
Same changes in the magnetic structure are generated by the displacement along the translational mode $\mathbf{Q}_\beta(\vec{r})$: 
\begin{equation}
\mathbf{m}_\beta(\vec{r}+\vec{e} dr)-\mathbf{m}_\beta(\vec{r})=\mathbf{Q}_\beta(\vec{r}) dq,
\end{equation}
where $dq$ is the magnitude of the displacement. 
The translational mode is proportional to $(\vec{e}\cdot \vec{\nabla}_r)\mathbf{m}_\beta(\vec{r})$:
\begin{equation}
\mathcal{N}_\beta\mathbf{Q}_\beta(\vec{r}) = (\vec{e}\cdot \vec{\nabla}_r)\mathbf{m}_\beta(\vec{r}),
\end{equation}
where factor $\mathcal{N}_\beta$ is fixed by the normalization condition, $|\mathbf{Q}_\beta(\vec{r})|=1$:
\begin{equation}
\mathcal{N}_\beta=\left|(\vec{e}\cdot \vec{\nabla}_r)\mathbf{m}_\beta(\vec{r})\right|.
\label{eq:s4}
\end{equation}

Equations~(\ref{eq:s1})-(\ref{eq:s4}) make it possible to replace the integration over the translational mode by integration over $r$: $dq=\mathcal{N}_\beta dr$. Specifically, integration over the two translational modes yields: 
\begin{equation}
V_\beta = \mathcal{N}_\beta^2 S,
\label{eq:s5}
\end{equation}
where $S$ is the area of the film. Note that $S$ cancels out in Eq.~(\ref{eq:prefactor}). 

In practice, normalization factor $\mathcal{N}_\beta$ can be obtained in various ways. Here, we simply used the finite-difference representation of the directional derivative in the RHS of Eq.~(\ref{eq:s4}) combined with the translation of the magnetic configuration by a lattice vector $\vec{a}$: 
\begin{equation}
\begin{split}
    \mathcal{N}_\beta &\approx \frac{1}{a}\left|\mathbf{m}_\beta(\vec{r}+\vec{a})-\mathbf{m}_\beta(\vec{r})\right|\\
    &=\frac{1}{a}\left(\sum_{j=1}^P\left|\vec{m}_{j,\beta}(\vec{r}+\vec{a})-\vec{m}_{j,\beta}(\vec{r})\right|^2\right)^{1/2}.
    \end{split}
\end{equation}

\section*{Data availability}
\noindent Data files containing results of intermediate calculations involved in the evaluation of the skyrmion lifetime for a particular value of magnetic interaction parameters are provided in the Data Set accompanying the article. All other data that support the findings of this study are available from the corresponding author upon reasonable request. 

\section*{Code availability}
\noindent The code used to calculate the results for this work is available from the authors upon reasonable request.

\section*{Acknowledgments}
\noindent The authors would like to thank V.M.~Uzdin, H.~J\'onsson, S.~Bl\"ugel, N.S.~Kiselev, G.P.~M\"uller, K.~von Bergmann, A.~Kubetzka, S.~Meyer, T.~Sigurj\'onsd\'ottir for helpful discussions and useful comments. 
This work was funded by the Russian Science Foundation (Grant No. 17-72-10195), the Icelandic Research Fund (Grants No. 163048-053, No. 185409-052 and No. 184949-052), the University of Iceland Research Fund, the European Union's Horizon 2020 Research and Innovation Programme under Grant Agreement No. 665095 (FET-Open project MAGicSky), and Alexander von Humboldt Foundation.
\section*{Competing financial interests}
\noindent The authors declare that they have no competing financial or non-financial interests. 

\section*{Author contributions}
\noindent A.S.V. and P.F.B. initiated the study. A.S.V. performed the calculations and prepared the figures. A.S.V, G.K. and S.v.M. analyzed the magnetic excitation spectra. A.S.V., S.v.M. and M.N.P. analyzed the pre-exponential factor. P.F.B. supervised the project. P.F.B. and S.H. wrote the manuscript. All of the authors discussed the results and contributed to the preparation of the article.
%


\end{document}